\documentclass[useAMS,usenatbib,a4paper]{mn2e}
\usepackage{graphicx}
\usepackage{amsmath}
\usepackage{txfonts}
\usepackage{mathrsfs}
\usepackage{color}

\def\aap{A\&A}
\def\apjl{ApJ}

\def\apjs{ApJS}

\newcommand{\be}{\begin{equation}}
\newcommand{\ee}{\end{equation}}
\newcommand{\bea}{\begin{eqnarray}}
\newcommand{\eea}{\end{eqnarray}}

\title[Alcock-Paczy\'nski Effect with SDSS-IV Quasars]{Assessing Cosmic Acceleration 
with the Alcock-Paczy\'nski Effect in the SDSS-IV Quasar Catalog}
\author[Melia, Qin \& Zhang]{Fulvio Melia$^{1}$\thanks{John Woodruff Simpson Fellow. 
E-mail: fmelia@email.arizona.edu},
{Jin Qin$^2$} and {Tong-Jie Zhang$^2$}\\
$^1$Department of Physics, The Applied Math Program, and Department of Astronomy, 
The University of Arizona, AZ 85721, USA\\
$^2$Department of Astronomy, Beijing Normal University, Beijing 100875, China}

\begin{document}

\date{}

\pagerange{\pageref{firstpage}--\pageref{lastpage}} \pubyear{2020}

\maketitle

\label{firstpage}

\begin{abstract}
The geometry of the Universe may be probed using the Alcock-Paczy\'nski
(AP) effect, in which the observed redshift size of a spherical distribution of
sources relative to its angular size varies according to the assumed cosmological model.
Past applications of this effect have been limited, however, by a paucity of suitable
sources and mitigating astrophysical factors, such as internal redshift-space
distortions and poorly known source evolution. In this {\it Letter}, we introduce
a new test based on the AP effect that avoids the use of spatially bound systems,
relying instead on sub-samples of quasars at redshifts $z\lesssim 1.5$ in the
Sloan Digital Sky Survey IV, with a possible extension to higher redshifts and improved
precision when this catalog is expanded by upcoming surveys. We here use this method
to probe the redshift-dependent expansion rate in three pertinent
Friedmann-Lema\^itre-Robertson-Walker (FLRW) cosmologies: $\Lambda$CDM,
which predicts a transition from deceleration to acceleration at $z\sim 0.7$;
Einstein-de Sitter, in which the Universe is always decelerating; and the
$R_{\rm h}=ct$ universe, which expands at a constant rate. $\Lambda$CDM is
consistent with these data, but $R_{\rm h}=ct$ is favoured overall.
\end{abstract}

\begin{keywords}
cosmological parameters, cosmology: observations, cosmology: theory, distance scale,
galaxies: active, quasars: supermassive black holes
\end{keywords}

\section{Introduction}
The expansion history of the Universe has been probed using a diverse set of observations, including those
of the Cosmic Microwave Background Radiation (CMB) and the large-scale structure of galaxy clusters. 
These approaches have been limited by processes other than those in the baseline cosmological
model, however, due to possible source evolution in the latter, or the generation and modification of 
anisotropies in the former \citep{Narlikar:2007,Angus:2011,Lopez:2013,Melia:2014,Lopez:2007}. Though 
large surveys of galaxies may constrain the geometry of the Universe, e.g., via the construction of 
a Hubble diagram or the implementation of an angular-size test, one must typically adopt specific 
astrophysical models, such as the growth of dark-matter halos, in order to extract useful cosmological 
information.

The geometry of the Universe can be assessed more cleanly via the Alcock-Paczy\'nski (AP)
\citep{Alcock:1979,Lopez:2014} effect, in which the ratio of observed radial/redshift size to
angular size of a spherical distribution of sources, such as a galaxy cluster, changes from
one cosmology to the next. AP tests largely avoid contamination from source evolution
because the characteristics of individual sources do not impact the ratio of projected
sizes of their distribution. Of course, to fully utilize the AP effect, one must have access
to bound systems that are large enough to be measurable over a broad range of redshifts, and
this tends to be a principal mitigating factor.

In this {\it Letter}, we introduce a new test based on the AP effect designed to probe the cosmic
expansion rate as a function of redshift, though using the very large sample of quasars in the
Sloan Digital Sky Survey IV (SDSS-IV) \citep{Ahumada:2020}, spanning redshifts $z\lesssim 3.5$. The novel
feature with this approach is that it does not rely on spatially confined source distributions. As
we shall see, the use of quasars can greatly improve the statistics for the purpose of model selection,
especially when future surveys will grow this catalog by two or more orders of magnitude.

\begin{figure*}
\centerline{\includegraphics[keepaspectratio,clip,width=6.5truein]{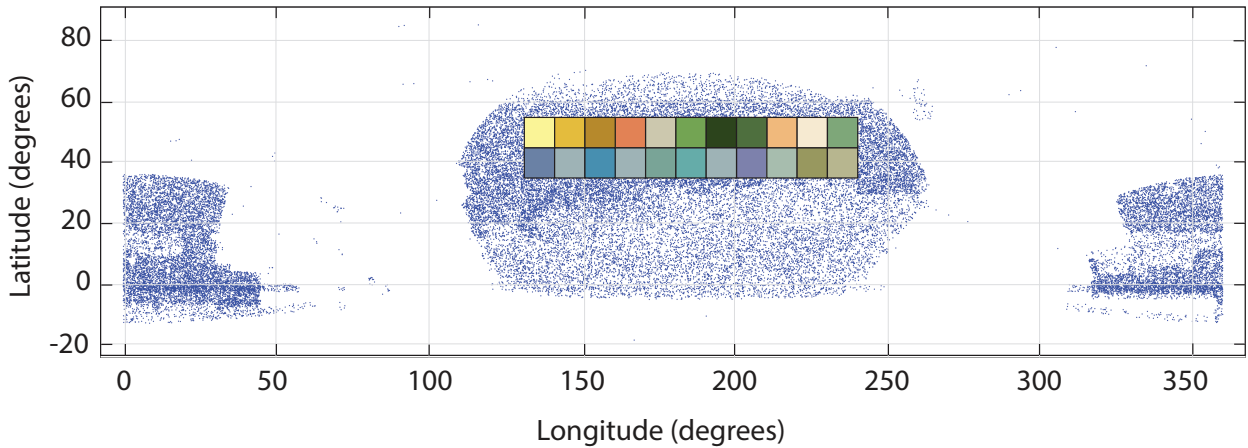}}
\caption{Sky map in Galactic coordinates $(l,b)$, of the projected locations (blue dots)
of the 526,356 SDSS-IV quasars. The colored boxes show sub-samples selected for the new AP test
(see text). For example, the yellow box has a size of $10^\circ\times 10^\circ$ and is centered
at Galactic coordinates $(135^\circ,55^\circ)$.}
\end{figure*}

Our modified AP test can be used for any cosmology but, given our focus on the impact of acceleration
on the geometry, we shall here restrict our attention to three
highly pertinent models: {\it Planck}-$\Lambda$CDM \citep{Planck:2018}, which predicts different
epochs of acceleration and deceleration; Einstein-de Sitter, which has been strongly ruled
out as a viable model of the Universe by many other kinds of data, but is included because
it provides a well-known example of a Universe that only decelerates; and another FLRW cosmology
based on the zero active mass equation-of-state, $\rho+3p=0$, in terms of the total energy
density $\rho$ and pressure $p$ in the cosmic fluid
\citep{Melia:2007,Melia:2016,Melia:2017a,Melia:2012,Melia:2020}. Known as the $R_{\rm h}=ct$
universe, this cosmology exhibits a constant rate of expansion throughtout its history.
(See Table~2 in ref.~\citealt{Melia:2018} for a more detailed comparison of this model with
$\Lambda$CDM.)

\vfill\newpage
\section{Data and Methodology}
All of the data we use in this {\it Letter} are taken from the Data Release 16 Quasar catalog
(DR16Q), based on the extended Baryon Oscillation Spectroscopic Survey (eBOSS) of the SDSS-IV
\citep{Ahumada:2020}. This collection includes all SDSS-IV/eBOSS objects spectroscopically confirmed
as quasars. With the inclusion of previously confirmed quasars from SDSS-I, II and III, this
combined catalog encompasses an overall sample of 526,356 objects, though possibly subject to
an estimated contamination of about $0.5\%$. This compilation spans a redshift range up to 
$\sim 3.5$, covering approximatey 9376 deg$^2$ of the sky (see fig.~1). As we shall see,
however, the quasar number density per comoving volume decreases non-negligibly at $z\gtrsim 1$
(see fig.~3), and we shall therefore restrict our attention to sources below this redshift for
this first application of the AP test. In addition, the quasar sample is less
complete at low latitudes (i.e., $b\lesssim 30^\circ$) compared to $b\gtrsim 30^\circ$, due 
in part to galactic foreground effects. As such, we shall here restrict our analysis to the 
sub-samples shown as colored boxes in figure~1 to maximize the statistics. 

Our methodology utilizes the geometric construction shown in figure~2. In short, we select
two adjacent {\it comoving} boxes, each with its four lateral sides inclined at a fixed angle
$\Delta\theta/2$ (in both galactic latitude $b$ and longitude $l$), relative to the line-of-sight
(LOS). We then count the total number of quasars, $N_{Q1}$, in box 1, based on its pre-selected
length, $\Delta z_1$, in redshift space (more on this below), and find from the quasar catalog
the value of $\Delta z_2$ for which $N_{Q2}=N_{Q1}$. For a fixed angular size
$\Delta\theta\times\Delta\theta$, the redshift interval $\Delta z_2$ is a unique function
of $\Delta z_1$ and the redshift-dependent comoving distance predicted by each given
cosmology. A comparison of the `measured' and theoretical relations between $\Delta z_1$ and
$\Delta z_2$ then provides a likelihood of that being the correct model.

\begin{figure}
\centerline{\includegraphics[keepaspectratio,clip,width=0.4\textwidth]{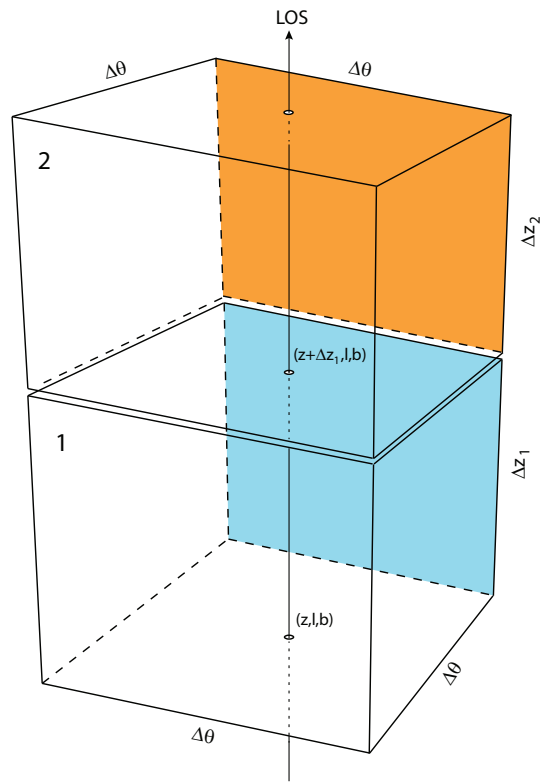}}
\caption{Schematic diagram showing two adjacent boxes along the line of sight
(LOS), each with angular dimensions $\Delta\theta\times \Delta\theta$ in the plane of
the sky. Box 1 has its base centered at $(z,l,b)$, and has a length $\Delta z_1$ in
redshift space. Box 2, whose base is centered at $(z+\Delta z_1,l,b)$, has the corresponding
length $\Delta z_2$.}
\end{figure}

In order for this method to provide a meaningful comparison, however, several conditions need to
be satisfied. Box 1 has its base centered at coordinates $(z,l,b)$, with an angular size
$\Delta\theta\times \Delta\theta$ in the plane of the sky. For convenience, we predict from
the chosen cosmological model the redshift size, $\Delta z_1$---or, equivalently, the angle
$\Delta\theta$---such that all three dimensions, which we shall call $L_\parallel(z)$ (in the 
radial direction) and $L_\perp(z)$ (in the two transverse directions at the base), are all 
equal in the comoving frame. As it turns out, the angular size $L_\perp(z)$ does 
not appear in the final analysis because $\Delta\theta$ remains constant throughout the region
$(z,z+\Delta z_1+\Delta z_2)$. $L_\perp(z)$ serves only to establish the area at the base of
the first box, but because $\Delta\theta$ remains constant, changing $L_\perp$ represents
a change in its area that is mirrored in proportion by the area of the second box. Thus,
the ratio $\Delta z_2/\Delta z_1$ is independent of $L_\perp(z)$, as one may see more formally
in the derived condition shown in Equation~(13) below. Choosing $\Delta\theta$ in this fashion 
(if $\Delta z_1$ is fixed) merey provides a convenient sub-sample of quasars with which
to calculate $N_{Q1}$ and $N_{Q2}$, and may be used for all the cosmologies being tested.

We make the key assumption that the {\it average} quasar comoving number density, $n_Q(z)$, within
the two boxes is uniform on a scale of several degrees. The average density is more likely to be
uniform across the boxes as their size increases, of course, though redshift evolution in $n_Q(z)$
could invalidate this approach if $dn_Q/dz$ from Box 1 to Box 2 is too large to ignore. Any potential
evolution in $n_Q$ may therefore be mitigated by choosing as small a box as possible, in order to
minimize the ratio $(dn_Q/dz)\Delta z/n_Q$. Unfortunately, these two requirements conflict each
other, but we have found by direct inspection of the SDSS-IV catalog that any dispersion arising
from these effects is well within the final calculated errors as long as $\Delta z_1/z\lesssim 1/3$
and $z\lesssim 1$. One may see this graphically in figure~3, which shows the estimated DR16Q quasar
number density per unit comoving volume, assuming {\it Planck}-$\Lambda$CDM as the background
cosmology (see Eqns.~5 and 6 below). (Note that this plot is merely used to gauge how reliable our
assumption of a constant $n_Q$ is, and is not included in the comparative analysis, which needs to
be carried out independently for each individual cosmology.)  Evidently, $n_Q$ is very nearly constant
up to $z\sim 1.2$, and then decreases monotonically towards higher redshifts.

\begin{figure}
\centerline{\includegraphics[keepaspectratio,clip,width=0.5\textwidth]{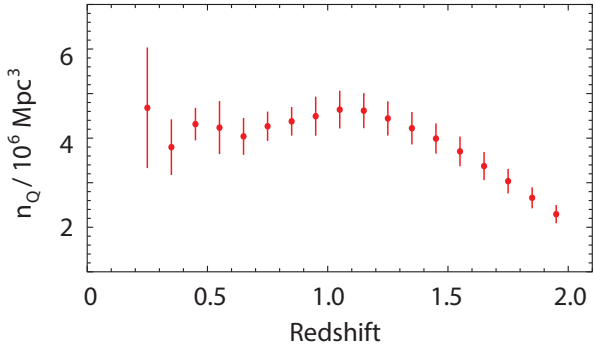}}
\caption{The DR16Q quasar number density as a function of redshift, assuming {\it Planck}-$\Lambda$CDM
as the background cosmology (see Eqns.~5 and 6).}
\end{figure}

Of the three cosmologies we shall test here, the geometry in $R_{\rm h}=ct$ is the
easiest to understand because all of its integrated measures---such as the comoving
distance---have simple analytical forms. We shall therefore start with this model, and
then summarize the corresponding expressions for the other two. For Box 1 along the LOS,
we have in the $R_{\rm h}=ct$ universe
\begin{equation}
L^{R_{\rm h}}_\parallel(z)={c\over H_0}\int_z^{z+\Delta z_1}{du\over E(u)}\;,
\end{equation}
where $H(z)\equiv H_0E(z)$, $E(z)=(1+z)$ and $H_0$ is the Hubble constant. For our actual calculations,
we shall employ the full integral expressions for these quantities. It will also be helpful, however,
for us to understand the results by finding approximations in the limit where $\Delta z\ll z$.
In this limit, Equation~(1) may be written
\begin{equation}
L^{R_{\rm h}}_\parallel(z)\approx {c\Delta z_1\over H_0(1+z)}\,.
\end{equation}
In the plane of the sky, the corresponding comoving size is
\begin{equation}
L^{R_{\rm h}}_\perp(z)\approx\Delta\theta {c\over H_0}\ln(1+z)\;.
\end{equation}
Thus, to estimate a reasonable size $\Delta z_1$ for Box 1 when $\Delta\theta$ is chosen, or
$\Delta\theta$ if $\Delta z_1$ is fixed, we may simply set these two expressions equal to each
other, $L^{R_{\rm h}}_\parallel(z)=L^{R_{\rm h}}_\perp(z)$, and find that
\begin{equation}
\Delta z_1\approx \Delta\theta (1+z)\ln(1+z)\;.
\end{equation}

For $\Lambda$CDM, the corresponding quantities are
\begin{equation}
L^{\Lambda}_\parallel(z)={c\over H_0}\int_z^{z+\Delta z_1}{du\over \sqrt{\Omega_{\rm m}(1+z)^3+\Omega_\Lambda}}\;,
\end{equation}
and
\begin{equation}
L^{\Lambda}_\perp(z)=\Delta\theta{c\over H_0}\int_0^z{du\over\sqrt{\Omega_{\rm m}(1+u)^3+\Omega_\Lambda}}\;.
\end{equation}
In these expressions, $\Omega_{\rm m}$ and $\Omega_\Lambda$ are today's matter and
dark-energy densities, respectively, scaled to the critical density $\rho_{\rm c}\equiv
3c^2H_0^2/8\pi G$. Radiation may be ignored for $z\lesssim 3.5$.
The parametrization shown in Equations~(5) and (6) is based on the {\it Planck}
optimization \citep{Planck:2018}: $\Omega_{\rm m}=0.315\pm0.007$, $\Omega_\Lambda=1.0-\Omega_{\rm m}$ (given
that $\Omega_k=0.001\pm 0.002$ is fully consistent with a perfectly flat Universe), and
an equation-of-state parameter $w_{\rm de}=-1.03\pm0.03$, where the pressure of dark energy is
written as $p_{\rm de}=w_{\rm de}\rho_{\rm de}$, in terms of its corresponding energy density
$\rho_{\rm de}$. This points to a cosmological constant, for which $p_{\rm de}= -\rho_{\rm de}$.
In our analysis, we shall therefore assume flat $\Lambda$CDM with $w_{\rm de}=-1$, though we shall
allow $\Omega_{\rm m}$ to vary in order to optimize the fit. This procedure is independent of all
other parameters, such as $H_0$. Finally, in the case of Einstein-de Sitter, we have
\begin{equation}
L^{\rm EdS}_\parallel(z)\approx {c\Delta z_1\over H_0(1+z)^{3/2}}
\end{equation}
and
\begin{equation}
L^{\rm EdS}_\perp(z)=\Delta\theta {2c\over H_0}\left(1-{1\over \sqrt{1+z}}\right)\,,
\end{equation}
which together yield
\begin{equation}
\Delta z_1\approx 2\Delta\theta(1+z)\left(\sqrt{1+z}-1\right)\,.
\end{equation}

\begin{figure}
\centerline{\includegraphics[keepaspectratio,clip,width=0.47\textwidth]{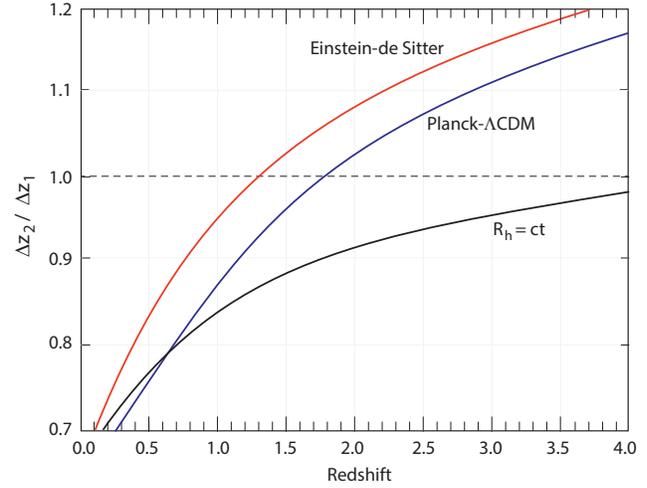}}
\caption{The ratio $\Delta z_2/\Delta z_1$ as a function of $z$, assuming $\Delta z_1=z/3$:
(blue) {\it Planck}-$\Lambda$CDM; (black) the $R_{\rm h}=ct$  universe; and (red) Einstein-de Sitter.
The standard model transitions from acceleration to deceleration across $z\sim 0.7$, while Einstein-de
Sitter always decelerates, and $R_{\rm h}=ct$ expands at a constant rate.}
\end{figure}

\begin{figure*}
\centerline{\includegraphics[keepaspectratio,clip,width=0.75\textwidth]{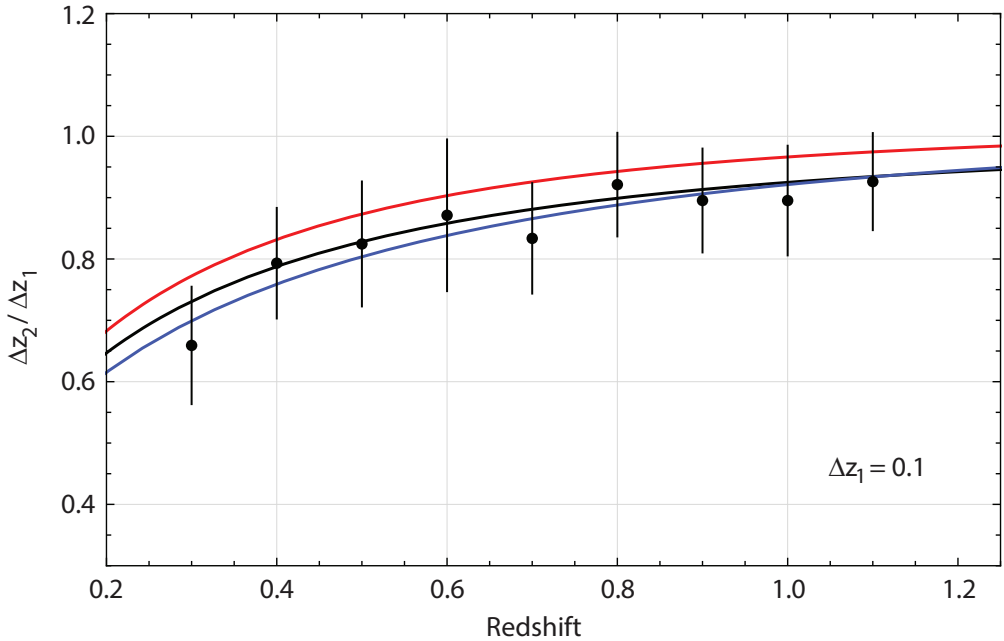}}
\caption{Same as figure~4, except now for a fixed $\Delta z_1=0.1$, highlighting the
variation with redshift at $0\lesssim z\lesssim 1.2$. The color coding is the same as in
figure~4, except that blue now corresponds to the best-fit $\Lambda$CDM model, with the
optimized matter density $\Omega_{\rm m}=0.22^{+0.21}_{-0.10}$. The data points are
obtained by counting quasars in the DR16Q catalog.}
\end{figure*}

Were we to ignore the $\delta z$-dependence of $L_\perp(z+\delta z)$ over 
the redshift range $\delta z\in (0,\Delta z_1)$ and $(\Delta z_1,\Delta z_1+\Delta z_2)$, 
the comoving volumes 1 and 2 would be equal if $\Delta z_2$ were chosen to satisfy the 
condition
\begin{equation}
\int_{z+\Delta z_1}^{z+\Delta z_1+\Delta z_2}{du\over E(z)}=
\int_{z}^{z+\Delta z_1}{du\over E(z)}\;.
\end{equation}
This is not a good approximation, however, even for boxes with $\Delta z_1/z\lesssim 1/3$. Fortunately,
it is quite straightforward to incorporate the redshift-dependence of $L_\perp$ into a calculation of the
comoving volume, via the inclusion of the expression
\begin{equation}
L_\perp(z+\delta z)=L_\parallel(z) 
\left[1+{I(z,z+\delta z)\over I(0,z)}\right]\;,
\end{equation}
where
\begin{equation}
I(z_1,z_2)\equiv\int_{z_1}^{z_2}{du\over E(u)}\;.
\end{equation}
Since $\Delta\theta$ remains constant across both boxes, it is not difficult to see that $\Delta z_2$
must satisfy the condition
\begin{equation}
I(0,z+\Delta z_1+\Delta z_2)^3=2I(0,z+\Delta z_1)^3-I(0,z)^3\;.
\end{equation}

The complete solution for $\Delta z_2/\Delta z_1$, taking all of these spherical effects into
account, is shown in figure~4 for each of the three cosmologies, using the criterion
$\Delta z_1=z/3$. We show in this figure the expected model differences over the extended range
($0\lesssim z \lesssim 4$), principally to demonstrate the importance of conducting this
test at high redshifts using the expanded quasar catalog that will be assembled with upcoming
surveys. Given the limitations of DR16Q (see fig.~3), however, we shall be focusing our attention
on the more restricted range ($0\lesssim z\lesssim 1$) in this paper (see fig.~5), 
as an illustration of the method.

It has been noted elsewhere (see, e.g., refs.~\citealt{Melia:2018,Melia:2020})
that the transition from deceleration to acceleration at $z\sim 0.7$ in $\Lambda$CDM produces
an overall expansion history very similar to that of $R_{\rm h}=ct$, at least up to $z\sim 1$.
The two phases effectively cancel out, producing an integrated expansion approximately equal
to what it would have been if the Universe had expanded at a constant rate from the beginning.
This is reflected in the overlap between the two $\Delta z_2/\Delta z_1$ (black and blue)
curves below $z\sim 1$ in figure~4. At higher redshifts ($z\gtrsim 1.5$), however, the
cosmological constant in Equations~(5) and (6) becomes relatively unimportant, and
$\Lambda$CDM exhibits the effects of deceleration analogous to Einstein-de Sitter.

To differentiate between cosmological models with different combinations of unknowns,
one typically uses information criteria, such as the Akaike Information Criterion,
${\rm AIC}\equiv\chi^2+2f$, where $f$ is the number of free parameters
\citep{Akaike:1974,Liddle:2007,Burnham:2002,Melia:2013}, and the Kullback Information Criterion,
${\rm KIC}\equiv\chi^2+3f$ \citep{Cavanaugh:2004}. A third variant, known as the
Bayes Information Criterion (\citealt{Schwarz:1978}), is defined as ${\rm BIC}=\chi^2+f\ln N$,
where $N$ is the number of data points. The BIC is actually not based on information theory, 
but rather on an asymptotic ($N\rightarrow\infty$) approximation to the outcome of a
conventional Bayesian inference procedure for deciding between models. It suppresses
overfitting much more than AIC and KIC if $N$ is large. In the application we are
considering here, $N\lesssim 30$, for which $\ln(30)\sim 3.4$, compared to the
analogous factor $2$ in the case of AIC, and $3$ for KIC. In this case, the BIC does 
not provide a perspective very different from the other two criteria, and there is no
point in including it in this {\it Letter}.

For model $\mathcal{M}_\alpha$, the
unnormalized confidence of it being `true' is the Akaike weight $\exp(-{\rm AIC}_\alpha/2)$.
Thus, its relative likelihood of being the correct choice is
\begin{equation}
P(\mathcal{M}_\alpha)= \frac{\exp(-{\rm AIC}_\alpha/2)}
{\sum_\beta\exp(-{\rm AIC}_\beta/2)}\;.
\end{equation}
\vskip0.1in\noindent
The various outcomes may also be characterized via the difference $\Delta {\rm AIC} \equiv 
{\rm AIC}_2\nobreak-{\rm AIC}_1$ (and similarly for KIC), which determines the extent to which
$\mathcal{M}_1$ is favoured over~$\mathcal{M}_2$. The result is judged `positive' when
$\Delta\sim 2-6$, `strong' for $\Delta\sim 6-10$, and `very strong' if $\Delta\gtrsim 10$.

\newpage
\section{Results and Discussion}
A direct comparison of these three curves with the `measurement' of $\Delta z_2/\Delta z_1$
using quasar counts in the DR16Q catalog is shown in figure~5. The errors associated with measuring
the position of each quasar, $(z,l,b)$, are insignificant compared to the dispersion in their
comoving number density $n_Q$. To estimate the errors shown here, we therefore assemble as many
boxes as possible within the same redshift bin $\Delta z_1$ (though clearly not at the same angular 
position), and then calculate the population variance from this distribution. The error bars shown
in figure~5 represent the $1\sigma$ uncertainties inferred from this variance.

\begin{table}
\begin{center}
{\footnotesize
\caption{Model Selection using AP with the SDSS-IV quasar catalog}
\begin{tabular}{lccccc}
\hline\hline
Model& $\chi^2$ & AIC & Prob & KIC & Prob \\
&&&(AIC)&&(KIC)\\
\hline
$R_{\rm h}=ct$  & $1.094$ & $1.095$ & $62.4\%$ & $1.095$  & $69.6\%$ \\
$\Lambda$CDM ($\Omega_{\rm m}=0.22^{+0.21}_{-0.10}$) & $0.827$ & $2.827$ & $26.3\%$ & $3.827$ & $17.8\%$ \\
Einstein-de Sitter  & $4.513$ & $4.513$ & $11.3\%$ & $4.513$ & $12.6\%$ \\
\hline\hline
\end{tabular}
}
\end{center}
\end{table}

The fits in this plot, and the summary of results given in Table~1, show that the data are
consistent with $\Lambda$CDM, but actually somewhat favour the $R_{\rm h}=ct$ cosmology, which
expands at a constant rate at all redshifts, while strongly rejecting Einstein-de Sitter. 
The optimized matter density $\Omega_{\rm m}=0.22^{+0.21}_{-0.10}$ for $\Lambda$CDM is fully 
consistent with the {\it Planck} measurement (i.e., $0.315\pm0.007$). In a head-to-head 
comparison, the result that $R_{\rm h}=ct$ is favoured over $\Lambda$CDM is judged positive, 
with an outcome $\Delta {\rm AIC}\sim 1.7$ and $\Delta {\rm KIC}\sim 2.7$, based on the DR16Q catalog. 

Unlike many other kinds of observation, these data are effectively independent of any model because
they are constrained by a fixed $\Delta\theta$ throughout the redshift range $(z,z+\Delta z_1)$
and $(z+\Delta z_1,z+\Delta z_1+\Delta z_2)$, and they do not depend on the actual value
of $L_\perp(z)$. Thus, the same set of data apply to all three curves in figure~5. For a fixed
$\Delta\theta$, however, the ratio $L_\parallel(z)/L_\perp(z)$ of the boxes does change in redshift
space according to each model's prediction of the angular diameter distance. As we may see from
this figure, the statistical weight of the DR16Q quasar catalog is already sufficient
for us to start distinguishing between the three models we have examined in this paper.

But clearly, the probative power of this technique will grow considerably when future surveys will
allow us to extend this test to redshifts as high as $\sim 3.5$ (see fig.~4). Upcoming campaigns
will expand the quasar catalog by several orders of magnitude. For example, projections for the Large
Synoptic Survey Telescope \citep{Izevic:2017} imply that perhaps as many as 50 million detections
will have been confirmed over the lifetime of this collaboration, enhancing the current
DR16Q catalog a hundredfold. With such improved statistics, one may even model $n_Q(z)$ individually
for each cosmology over redshifts where the comoving density is not constant (see fig.~3), thereby
attaining an even higher level of precision for the $\Delta z_2/\Delta z_1$ curves.

\section*{Acknowledgments} 
We are grateful to the anonymous referee for several suggestions to improve
the presention in our manuscript. FM is also grateful to Amherst College for its support 
through a John Woodruff Simpson Lectureship. This work was partially supported by the National 
Science Foundation of China (Grants No.11929301,11573006) and the National Key R \& D Program 
of China (2017YFA0402600).

\section*{Data Availability Statement}
All of the data underlying this paper are taken from the Data Release 14 Quasar catalog
(DR16Q), based on the extended Baryon Oscillation Spectroscopic Survey (eBOSS) of the
SDSS-IV, which is fully described in \cite{Ahumada:2020}. The DR16Q catalog may be 
downloaded in fits format from the website https://www.sdss.org/dr16/.

\bibliographystyle{apj}
\bibliography{ms}

\label{lastpage}

\end{document}